\documentclass[12pt]{iopart}
\usepackage{iopams}
\usepackage{graphicx}
\begin{document}

\title{Noncommutative geometry inspired dirty black holes}

\author{P Nicolini$^1$\footnote{Current Address: Institut f\"{u}r Theoretische Physik, Johann Wolfgang Goethe Universit\"{a}t, Max von Laue Str.1, Frankfurt am Main, 60438 Germany.} and E Spallucci$^2$}

\address{$^1$Physics Department,
CSU Fresno, Fresno, CA 93740-8031, USA and Dipartimento di Matematica e Informatica, Consorzio di Magnetofluidodinamica,
Universit\`a degli Studi di Trieste and I.N.F.N., Via Valerio 12, 34127 Trieste, Italy
\\
$^2$Dipartimento di Fisica, Universit\`a degli Studi di Trieste
and I.N.F.N., Strada~Costiera 11, 34014 Trieste, Italy}

\ead{\mailto{nicolini@th.physik.uni-frankfurt.de}, \mailto{spallucci@trieste.infn.it}}

\begin{abstract}
We provide a new exact solution of the Einstein equations which generalizes the noncommutative
geometry inspired Schwarzschild metric, we previously obtained. We consider here  more
general relations between the energy density and the radial pressure and find  a new geometry describing  a regular
``dirty black hole''. We discuss strong and weak energy condition violations and
various aspects of the regular dirty black hole thermodynamics.

\end{abstract}

\maketitle

\section{Introduction}
In a recent series of papers we obtained exact solutions of the Einstein equations describing
both neutral and charged black holes free of curvature singularities in the origin
\cite{Nicolini06,Rizzo06,Ansoldi07,Spallucci:2008ez,Casadio08,Ansoldi08,Nicolini08,Arraut:2009an}.
We reached these conclusions after a long path starting from an original approach to
noncommutative geometry which is based on coordinate coherent states
\cite{Smailagic:2001qe,Smailagic:2002sp,Smailagic:2003yb,Smailagic:2003rp,Smailagic:2004yy,Nicolini:2005de,Nicolini05,Spallucci06,Banerjee:2009gr,Nicolini:2009dr}, in alternative to the mathematically correct, but physically hard to implement,  ``star-product'' formulation.\\
However, our results are ``model-independent'' in the sense that our approach improve the short-distance
behaviour of the Einstein equations by taking into account the presence of a quantum gravity
induced minimal length, whatever it is.

String Theory, Noncommutative Geometry, Generalized Uncertainty Principle, etc., all point out
the existence of a lower bound to distance measurements.  Thus, the very concept
of ``point-like'' particle becomes physically meaningless and must be replaced with its
best approximation consistent with the tenets of quantum mechanics, i.e. a minimal width
Gaussian distribution of mass/energy.  Solving the Einstein equations for a static, minimal width,
mass-energy distribution centered around the origin, we found black hole type solution smoothly
interpolating between deSitter spacetime at short distance and Schwarzschild geometry at large
distance. The characteristic length scale of this system is given by the matter distribution width $\sqrt\theta$.\\
The geometric and thermodynamics features of the solution can be summarized as follows:\\
i) there is no curvature singularity in $r=0$; the center of the black hole is a regular ``ball'' of
deSitter vacuum accounting for the short-distance quantum fluctuations of the spacetime manifold,
whatever is their physical origin;\\
ii) even in the neutral case, there exist an outer and an inner horizon. At the end of the
Hawking evaporation the two horizons coalesce into a single, degenerate horizon, corresponding
to an extremal black hole;\\
iii) the Hawking temperature reaches a finite maximum value and the drops down to zero
     in the extremal configuration.\\
To achieve this regular behaviour the choice of the matter equation of state is instrumental.
Stability of the solution is guaranteed by choosing

\begin{equation}
 \rho(r)= -p_r(r)
\label{vacuum}
\end{equation}
where, $ \rho(r)$, $p_r(r) $ are the Gaussian matter density and radial pressure, respectively.
The tangential pressure $ p_\perp(r)$ was determined by   $\rho(r)$  and $p_r(r)$ through
hydrodynamic equilibrium equation.
The condition (\ref{vacuum}) has  the same form of the vacuum equation of state. Thus, one
can expect that near the origin the metric will be deSitter with an effective cosmological
constant $\Lambda \propto G_N \,\rho(0) $.  On the other hand, if the width of the
matter distribution is $\sqrt\theta$, at large distance one sees a small sphere of matter
with radius about  $\sqrt\theta$. Thus, Birkhoff theorem assures the metric to be Schwarzschild.
In the intermediate region the metric is neither deSitter, nor Schwarzschild, and can be
analytically written in terms of lower incomplete gamma function.

In this paper we are going to relax equation (\ref{vacuum}) and look for a new  solution
describing a dirty black hole.
The starting point is the minimal width, Gaussian, mass/energy distribution

\begin{equation}
 \rho\left(\, r\,\right)=  \frac{M}{\left(\, 4\pi\theta\, \right)^{3/2}} \exp{\left(\,
 -\frac{r^2}{4\theta}\,\right)}
\label{gdistr}
\end{equation}
which we obtained from our coordinate coherent states approach to noncommutative geometry. From this
perspective the $\theta$ parameter is a length squared quantity defining the scale where spacetime
coordinates become non-commuting (quantum) objects. However, in a more general framework the
distribution (\ref{gdistr}) represents the most localized energy density which is compatible
with the existence of a minimal length, whatever it is. Thus, the width of the bell-shaped
function (\ref{gdistr}) can be consistently related to the string length $\sqrt{\alpha^\prime}$,
the TeV Quantum Gravity scale, etc. In a recent paper \cite{Sushkov:2005kj} a gaussian source has
been used to model phantom energy supported wormholes.\\
The constant $M$ is the total mass energy given by

\begin{equation}
M\equiv 4\pi \int^\infty_0 dr r^2 \rho\left(\, r\,\right).
\end{equation}
The mass distribution (\ref{gdistr}) is the component $T^0{}_0$ of the energy momentum tensor.
Before proceeding we have to define the remaining components. We model our source through
a fluid-type $T^\mu{}_\nu$ of the following form:

\begin{equation}
 T^\mu{}_\nu=\mathrm{Diag}\left(\, -\rho\left(\, r\,\right)\ , p_r\left(\, r\,\right)\ ,
 p_\perp\left(\, r\,\right)\ , p_\perp\left(\, r\,\right)\,\right)
 \end{equation}
where, $ p_r$ is  the \textit{radial} pressure and $ p_\perp$ the \textit{tangential} pressure.\\
In order to be a viable source for the Einstein equations, the condition $T^\mu{}_\nu\ ;\mu=0$ must be satisfied
by $\rho$, $p_r$, $p_\perp$.

We are going to solve Einstein equations assuming the metric to be spherically symmetric,
static, asymptotically flat. Thus, we can write the general form of the line element
in terms of two independent functions $\Phi\left(\, r\,\right)$ and $m\left(\, r\,\right)$ as

\begin{equation}
ds^2=-e^{2\Phi\left(\,r\,\right)}dt^2+
\frac{dr^2}{ 1-2m\left(\,r\,\right)/r}
+r^2 \,\left(\, d\psi^2+\sin^2\psi \, d\phi^2\,\right)
\label{metric}
\end{equation}
where $m\left(\,r\,\right)$ is the  \textit{shape function} and
$\Phi\left(\,r\,\right)$  is the  \textit{red-shift function} \cite{Morris:1988cz,Visser:1992qh}. Both unknown functions
must be determined by the Einstein equations.
To this purpose, we now briefly recall a recent solution one obtains by assuming in place of (\ref{vacuum}) the condition

\begin{equation}
p_r=-\frac{1}{4\pi r^3\, }m\left(\,r\,\right).
\label{eqstato}
\end{equation}
where

\begin{equation}
m\left(\,r\,\right)\equiv 4\pi\int_0^r du\,u^2\,\rho\left(\,u\,\right).
\end{equation}

Thus, the resulting line element reads

\begin{equation}
ds^2=-dt^2+\frac{dr^2}{1-4M\gamma\left(\,3/2\ ; r^2/4\theta\,\right)/\sqrt{\pi} r} +r^2 \left(\, d\vartheta^2+\sin^2 \vartheta \, d\phi^2\,\right)
\label{w}
\end{equation}

which describes a wormhole geometry. Its properties
have been recently investigated in \cite{Garattini:2008xz} and we are not going to repeat them here. We have recalled this wormhole solution since it turns out to be useful to understand the procedure we shall follow in the next section. Indeed to derive the dirty black hole solution, we are going to relax the condition (\ref{eqstato}) in favor of a more general one motivated by physical requirements.

\section{Dirty Black Hole}

The term ``dirty black holes'' refers to black hole solutions of the Einstein
equations in interaction with various kind of matter fields\footnote{This kind of solutions are alternatively
referred to as ``\textit{hairy}'' black holes. }, some remarkable examples
are:
\begin{itemize}
 \item gravity + electromagnetism + dilaton \cite{Gibbons:1987ps,Ichinose:1989vb,Yamazaki:1992pm,Garfinkle:1990qj};
 \item gravity + electromagnetism + axion \cite{Allen:1989kc,Campbell:1991rz,Lee:1991jw};
 \item gravity + electromagnetism + Abelian Higgs field \cite{Dowker:1991qe};
 \item gravity + electromagnetism + dilaton+ axion \cite{Shapere:1991ta};
 \item gravity + non-Abelian gauge fields \cite{Galtsov:1989ip,Straumann:1990as,Bizon:1990sr,Bizon:1991nt};
 \item gravity +axion + non-Abelian gauge fields \cite{Lahiri:1992yz};
 \end{itemize}

In our model we do not select any specific kind of field theory, but we simply introduce
a  smeared energy/pressure  distribution, which is a classical parametrization for
the energy/pressure ``stored'' in some kind of field.
However, the condition which provides the relation between energy and radial pressure is crucial to determine the new solution. Among the huge variety of possible choices of this condition, we require that the pressures $p_r$ and $p_\perp$ satisfy the following physical conditions:
\begin{enumerate}
\item $p_r$ and $p_\perp$ must be asymptotically vanishing; \label{uno}
\item $p_r$ and $p_\perp$ must be finite at the horizon(s);\label{due}
\item $p_r$ and $p_\perp$ must be finite at the origin.\label{tre}
\end{enumerate}
 The condition \ref{uno}. assures that the solution matches the Minkowski space at infinity while the conditions \ref{due}. and \ref{tre}. warrant the regularity at the horizon(s) and at the origin, a fact that is in the spirit of all the previous solutions generated by a smeared source term.

The line element we are looking for is of the form:

\begin{equation}
\fl
ds^2=-e^{2\Phi\left(\,r\,\right)}\left(\, 1 -2m\left(\,r\,\right)/r \,\right)\, dt^2+
\frac{dr^2}{ 1-2m\left(\,r\,\right)/r}
+r^2 \,\left(\, d\psi^2+\sin^2\psi \, d\phi^2\,\right)
\label{dirtybh}
\end{equation}
which can be obtained form (\ref{metric}) by the following local rescaling of the red-shift function
$\Phi\left(\, r\,\right)\longrightarrow \Phi\left(\, r\,\right)+\frac{1}{2}\ln\left(\, 1-2m/r\,\right)$.
Therefore the Einstein equation read

\begin{eqnarray}
&&\frac{dm}{dr}= 4\pi\, r^2\rho\ ,\label{eq1}\\
&&\frac{1}{2g_{00}}\frac{dg_{00}}{dr}=\frac{m(r)+4\pi \ r^3 p_r}{r(r-2m(r))}\ , \label{eq22}\\
&&\frac{dp_r}{dr}= - \frac{1}{2g_{00}}\frac{dg_{00}}{dr}\left(\, \rho+p_r\,\right)+\frac{2}{r}\left(\, p_\perp-p_r \,\right)
\label{eq33}
\end{eqnarray}
where, $g_{00}\equiv  -e^{2\Phi\left(\,r\,\right)}\left(\, 1 -2m\left(\,r\,\right)/r \,\right)$.\\

 According to the recipes \ref{uno}., \ref{due} and \ref{tre}, pressures regularity requirements are satisfied by the following condition replacing the (\ref{eqstato})

\begin{equation}
\rho (r) + p_r(r)\equiv -\sqrt{\theta}\left(\, 1 -2m/r \right)
\frac{d\rho}{dr}=  \frac{1}{2\sqrt{\theta}}\, r \, \rho \, \left(\, 1 -2m/r \right)
\label{steqn}
\end{equation}
that can be rewritten as

\begin{equation}
p_r(r)=-\rho (r)\left[\,1-\frac{r}{2\sqrt{\theta}}\, \left(\, 1 -2m/r \right)\,\right] .
\label{steqn2}
\end{equation}
This is the simplest choice, satisfying the physical conditions above. In principle, one can still consider further conditions, involving additional terms $\sim (r/\sqrt{\theta})^m  (\, 1 -2m/r )^n$ in the square brackets on the r.h.s.. In such a case, one ends up with a more complicated solution, endowed only with subleading contributions to the solution we are going to derive, in the three physically meaningful regions (infinity, horizon(s) and origin).

From (\ref{eq33}) one finds that the angular pressure $p_\perp$ is

\begin{equation}
p_\perp (r)=p_r(r) +\frac{r}{2}\
\frac{dp_r}{dr}+\frac{1}{4\sqrt{\theta}}\ \rho(r)\ \left( m(r)+4\pi \ r^3 p_r \right).
\label{steqn3}
\end{equation}
We notice that both $p_r$ and $p_\perp$ enjoy the above physical conditions at the origin, horizon(s) and asymptotically. From Eqs. (\ref{eq22}) and (\ref{steqn2}) one has

\begin{equation}
 \frac{d\Phi}{dr}=\frac{4\pi r (\rho + p_r)}{1-2m/r}=\frac{2\pi r^2}{\sqrt{\theta}}\ \rho(r)
\end{equation}
with the boundary condition $ \Phi\left(\infty\right)=0$ in order to reproduce Minkowski geometry at infinity.
The solution reads

\begin{equation}
\Phi(r)= \frac{MG}{\sqrt{\pi\theta}}\gamma\left(\,3/2\ , r^2/4\theta\,\right) -\frac{MG}{2\sqrt\theta}
\end{equation}
where, for the sake of clarity we momentarily re-introduced the Newton constant $G$.

As a result the noncommutative Schwarzschild ``dirty'' black hole is described by the following line element

\begin{eqnarray}
\fl
 ds^2=-\exp{\left[\, -\frac{MG}{\sqrt\theta} \left(\,1 -\frac{2}{\sqrt\pi}  \gamma\left(\,3/2\ , r^2/4\theta\,\right)  \,\right)\,\right]}\left(1-2G m(r)/r\right) \, dt^2 + \frac{dr^2}{ 1-2G m(r)/r} \nonumber
\\ + r^2\, d\Omega^2
\end{eqnarray}
Before discussing thermodynamical properties, let us check the regularity of the geometry
at the origin. From the Einstein equations one finds

\begin{equation}
R(0)=-8\pi G T^\mu_\mu(0)=8\pi G\left[\, \rho(0) - p_r(0)- 2p_\perp(0)\,\right].
\end{equation}
From (\ref{steqn}), (\ref{steqn2}), (\ref{steqn3}) we see that all the pressures
approach $-\rho(0)$ near the origin. Thus, we find

\begin{equation}
R(0)=32\pi G\, \rho(0)=\frac{4GM}{\sqrt\pi\theta^{3/2}}\equiv 4\Lambda_{eff}
\end{equation}
which is the Ricci scalar for a deSitter metric characterized by a positive effective cosmological
constant $\Lambda_{eff} $. The same conclusion can be obtained by expanding the
line element near $r=0$:

\begin{equation}
\fl
ds^2 = -e^{-M/\sqrt/\theta}\left(\, 1 -\frac{\Lambda_{eff}}{3}r^2\,\right) dt^2
+\left(\, 1 -\frac{\Lambda_{eff}}{3}r^2\,\right)^{-1} dr^2 +r^2
\left(\, d\psi^2+\sin^2\psi \ d\phi^2\,\right)\label{desitter}
\end{equation}
where, the constant $e^{-M/\sqrt\theta}$ can be reabsorbed into a rescaled time coordinate
in order to have the ordinary deSitter line element.
Also in the case of the dirty black hole there is no curvature singularity and the
central black hole geometry results to be a deSitter spacetime.

Regarding the horizon(s), we have to study the equation $ M=U\left(\, r_H\,\right)$
where the ``effective potential'' $U\left(\, r_H\,\right)$ is defined to be

\begin{equation}
U\left(\, r_H\,\right)=
\frac{4}{\sqrt\pi} \frac{r_H}{\gamma\left(\,3/2\ , r^2_H/4\theta\,\right) }.
\label{eqhor}
\end{equation}
For a positive given value of $M$, the (\ref{eqhor}) can admit two, one, or no solution.
We studied in detail the solutions of this equation in \cite{Nicolini06} and we are not going to repeat all the discussion here.
It may be useful to summarize the main conclusion, i.e. the existence of a lower bound for a black hole
mass given by  $M_0\approx 1.9 \sqrt{\theta}/G$.
Thus, we find again three possible cases:
\begin{itemize}
 \item $M>M_0$ non-extremal dirty black hole with two horizons $r_+>r_-$; the outer horizon $r_+$ is said to be of {\it canonical type}, while the $r_-$ is a Cauchy
horizon.
 \item $M=M_0$ extremal dirty black hole with one degenerate horizon $r_+=r_-\equiv r_0\approx 3\sqrt{\theta}$; the function $g_{rr}$ does not change sign at the horizon, which is
said to be of {\it non-canonical type}.
 \item $0<M< M_0$ dirty minigravastar, no horizons.
\end{itemize}
\begin{figure}[ht!]
\begin{center}
\includegraphics[width=8.5cm,angle=0]{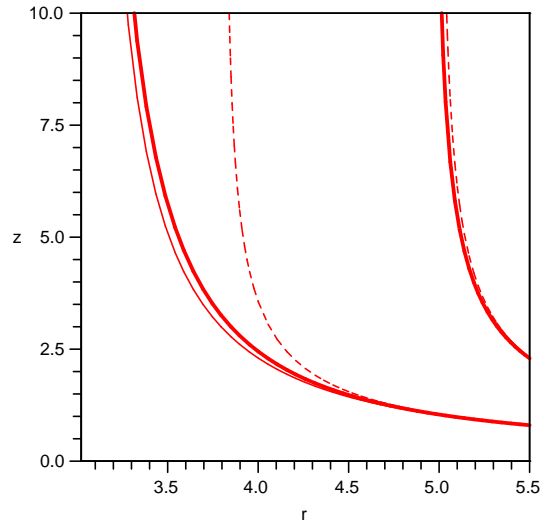}
\caption{\label{redbh}The asymptotic redshift $z=\Delta \lambda/\lambda$ versus the radius $r/\sqrt{\theta}$. On the right part of the figure, the curves are for the mass
 $M=2.5\sqrt{\theta}/G$, while on the left are for $M=M_0$.  The dashed curves corresponds to Schwarzschild case, while the solid ones are for the noncommutative
Schwarzschild (thin) and the noncommutative dirty case (thick). We can observe that, for $M=2.5\sqrt{\theta}/G$ all the curves coincide and we can conclude that
noncommutative effects are not yet important, while for smaller masses in the vicinity of the extremal configuration the curves become distinct.
We can conclude that the shape function $m(r)$ lowers the redshif, as in the case of
 a smaller gravitational field, while the redshif function $\Phi(r)$ slightly increases the values of $z$. }
\end{center}
\end{figure}

\begin{figure}[ht!]
\begin{center}
\includegraphics[width=8.5cm,angle=0]{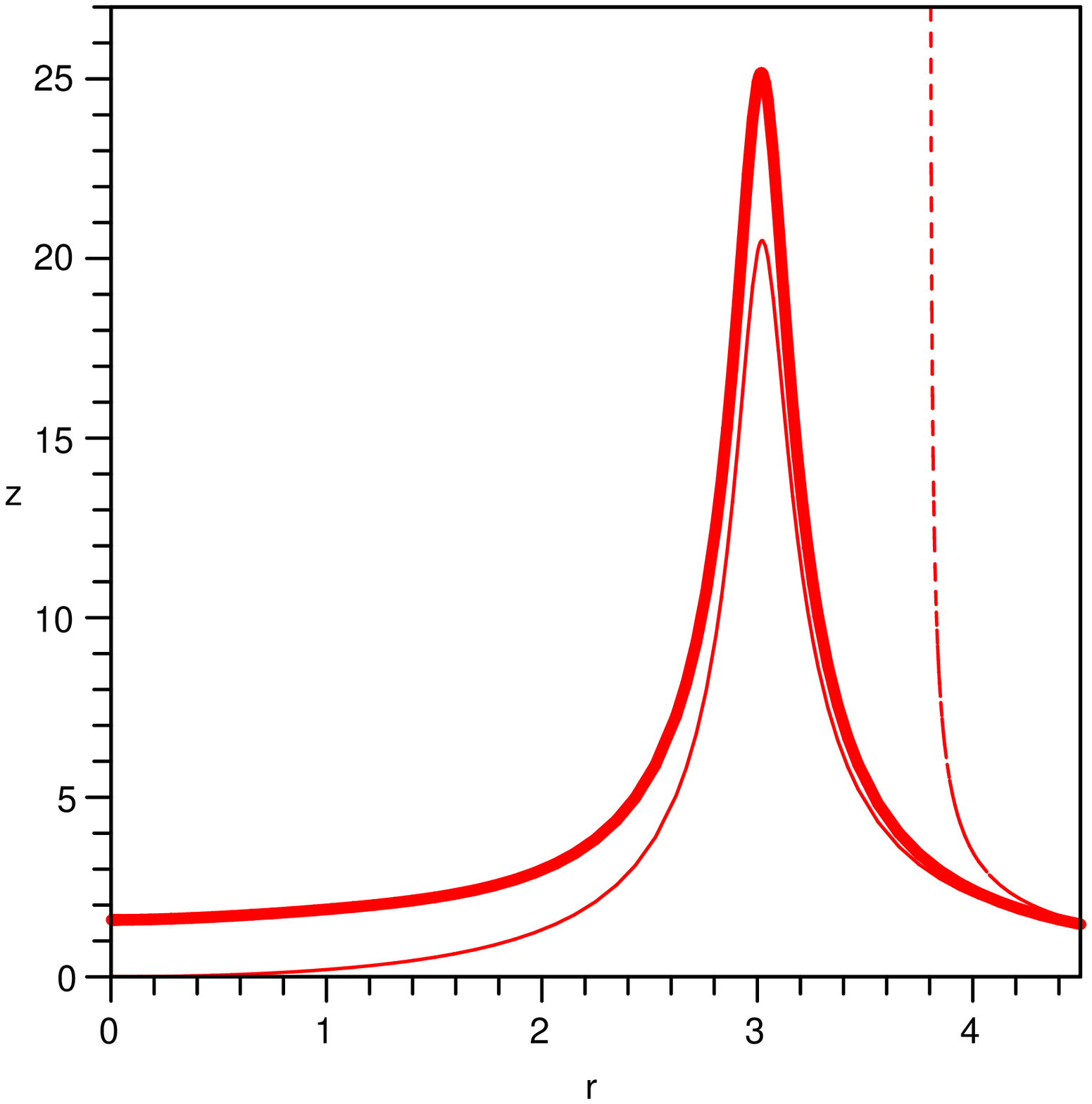}
\caption{\label{redgrava} The asymptotic  redshift $z=\Delta \lambda/\lambda$ versus the radius $r/\sqrt{\theta}$ for the minigravastar case, i.e. $M<M_0$.
The dashed curve correspond to the Schwarzschild line element and exhibit a divergent behavior due to the presence of horizons.
The thin solid curve corresponds to the noncommutative Schwarzschild case, while the thick solid curve is for the noncommutative dirty case. As a general
prescription, the shape function $m(r)$ provides a regular peak at $r=r_0$, while the redshif function $\Phi(r)$ slightly increases the values of $z$ with respect the
noncommutative Schwarzshild case for all $r$.}
\end{center}
\end{figure}

An interesting feature of the above line element regards the gravitational redshift
$z\equiv \Delta \lambda/\lambda$, where $\lambda$ is is the wavelength of the
electromagnetic radiation at the source and $\Delta \lambda$ is the difference between the observed and
 emitted wavelengths. This quantity now depends both on the
redshift function $\Phi(r)$ and the shape function $m(r)$.
The redshift measured by an asymptotic observer turns out to be

\begin{equation}
z=e^{\frac{MG}{2\sqrt\theta} \left(\,1 -\frac{2}{\sqrt\pi}  \gamma\left(\,3/2\ , r^2/4\theta\,\right)  \,\right)}\, \left(1-2G m(r)/r\right)^{-1/2}-1
\end{equation}
which is a clearly divergent quantity approaching the horizon, while asymptotically vanishes.
On the other hand, in the absence of horizons i.e. the minigravastar case, the
redshift $z$ is finite even at the origin.
\begin{figure}[ht!]
\begin{center}
\includegraphics[width=8.5cm,angle=0]{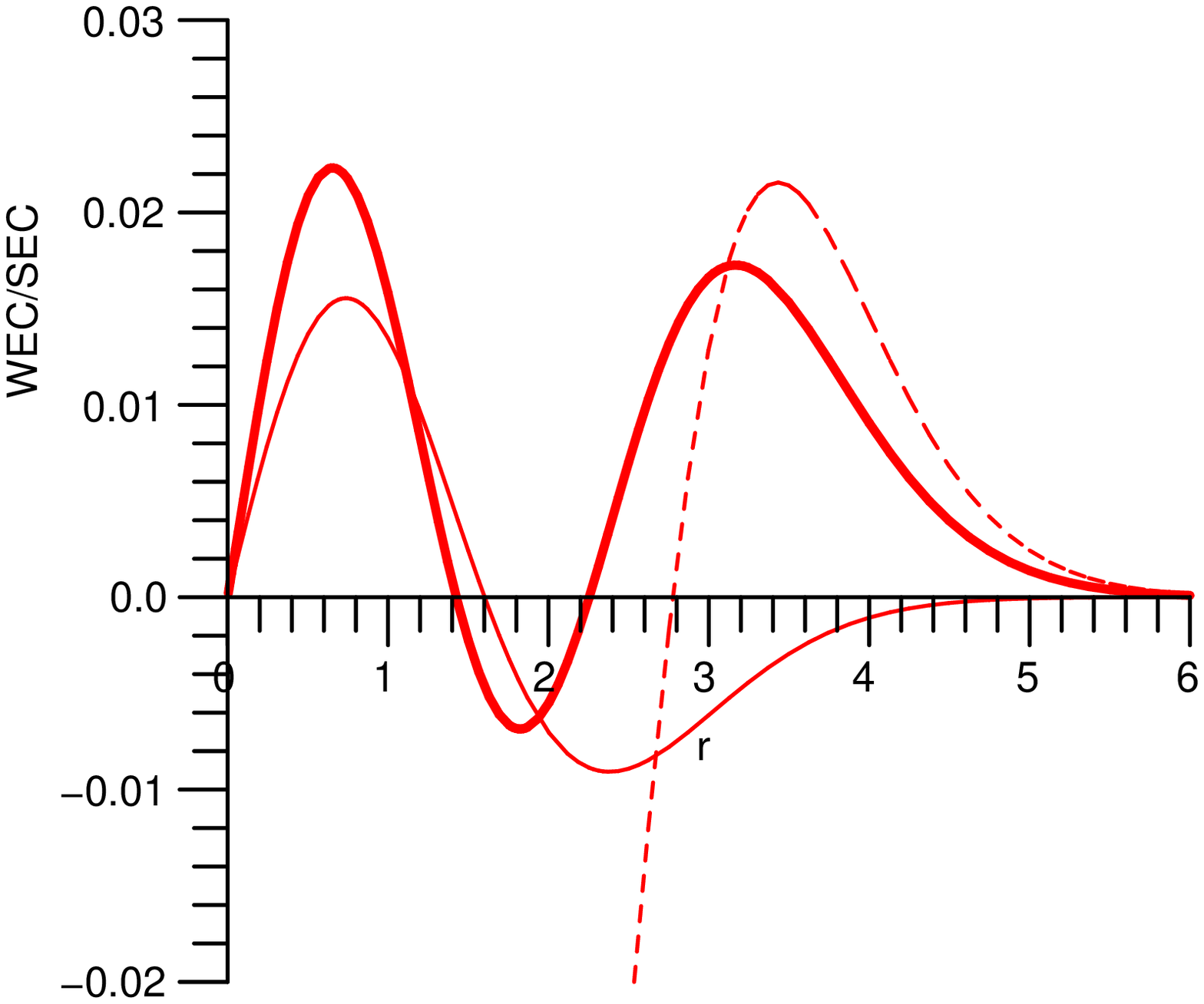}
\caption{\label{WSEC}The energy conditions for  $M=3.0\sqrt{\theta}/G$ versus $r/\sqrt{\theta}$. The thick solid curve is for the functions  $\rho + p_\perp$, while the thin one
corresponds to the function  $\rho + p_r$, which vanishes at the horizons.  The violation of the weak energy condition occurs when either one of
the solid curves have negative values. The dashed curve is the function $\rho + p_r +2 p_\perp$, whose negative values take place where the strong energy condition
is violated.}
\end{center}
\end{figure}

In the framework of ``wormhole engineering'' exotic matter is advocated in order to violate energy conditions and avoid the tunnel to collapse. Thus, weak (WEC) and/or strong
energy (SEC) conditions violation are an important issue to discuss in the black hole case, as well. Indeed, an eventual violation would mark a substantial departure from the behaviour of any kind of classical matter. This is what we expect as the Gaussian distribution, which is the corner stone of our approach and was originally derived  in \cite{Smailagic:2003rp} in a \textit{quantum} framework. The figure (\ref{WSEC}) shows the new features of the dirty black hole. \\
Let us start from the dashed curve representing the plot of the function  $\rho + p_r +2 p_\perp$. Where it is negative,
the SEC is violated. This occurs well for $r< 2.8\sqrt{\theta}$, which is well inside the event horizon located at $r_+ \approx 5\sqrt{\theta}$, as in the case of the ``clean'' noncommutative geometry inspired Schwarzschild black hole. Both in the case of dirty and clean solutions, the SEC is a short-distance effect caused by
the underlying non-commutative geometry.
Spacetime fluctuations provide an effective gravitational repulsion smearing out the classical curvature singularity.
On the other hand, outside the black hole matter behaves in a ``~classical~'' manner. However, the memory
of the short-distance effects is still present in the black hole temperature as it will be shown below.
The new distinctive feature of the dirty black hole is that WEC are violated as well, while
in the clean case they are preserved. Violation occurs, once again, inside the horizon and more
specifically where the two solid curves in figure (\ref{WSEC}) lower below the r-axis.
 The presence of some ``dirt''
 on the event horizon causes a suppression of the Hawking temperature with respect to the
corresponding clean case, in agreement with the general result shown in  {\cite{{Visser:1992qh}}}.

\begin{figure}
\begin{center}
\includegraphics[width=8.5cm,angle=0]{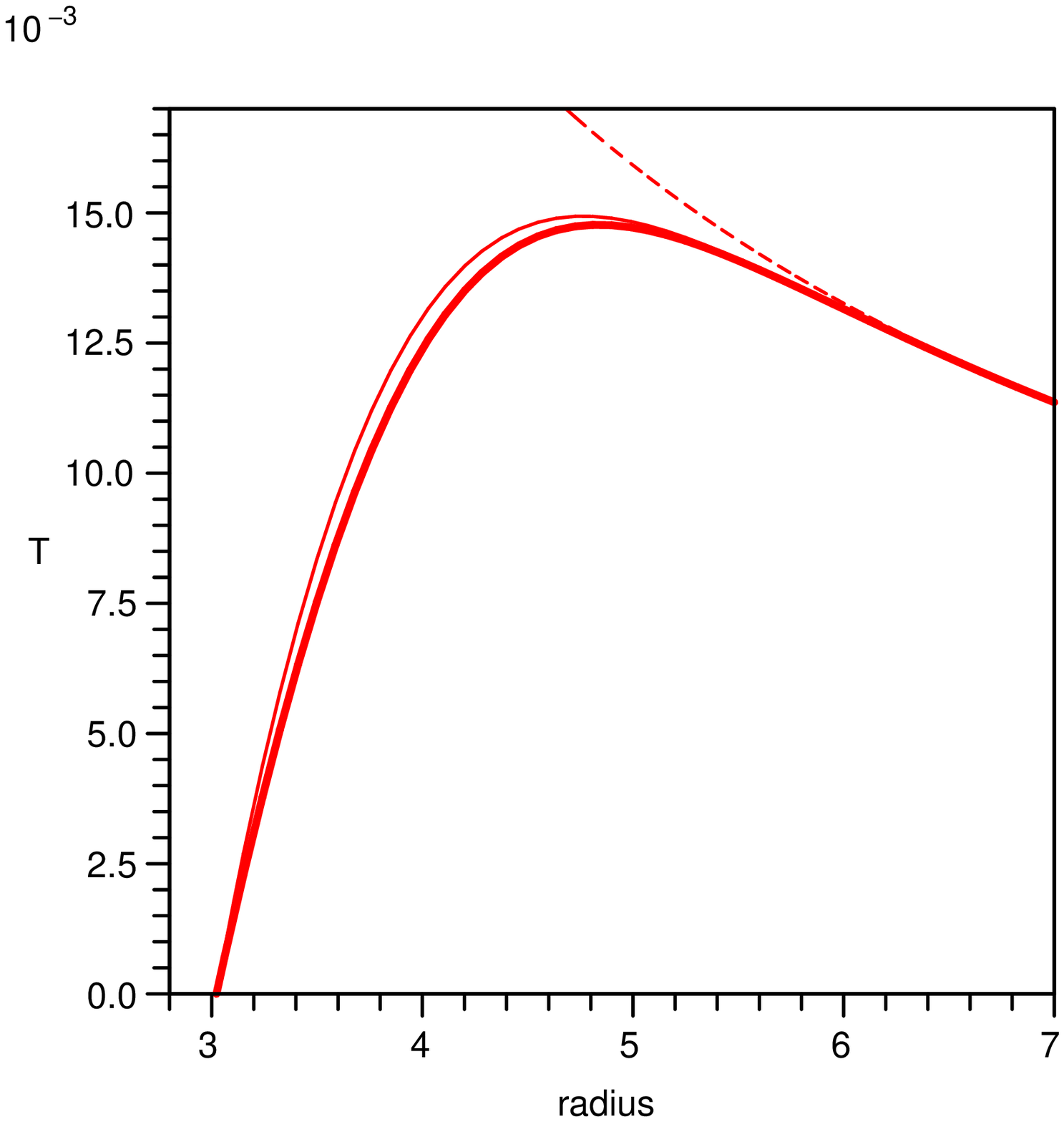}
\caption{\label{temp} The Hawking temperature $T_H$ versus the horizon radius $r_+$. The dashed curve corresponds
to Schwarzschild temperature, while from top to bottom the solid curves are for the noncommutative Schwarzschild (thin)
and the noncommutative dirty case (thick). We can observe that, while the temperature decreases for the dirty case, the
remnant radius is unchanged.}
\end{center}
\end{figure}

The Hawking temperature, mentioned above, is given by ($G=1\ , \kappa_B=1$):

\begin{eqnarray}
&& T_H=\frac{1}{4\pi r_+}\, e^{\Phi\left(\, r_+\,\right)}\, \left[1-\frac{r_+^3}{4\theta^{3/2}}\frac{e^{-r_+^2/4\theta}}{\gamma(3/2; r_+^2/4\theta)}\right]
\ ,\\
&& \Phi\left(\, r_+\,\right)\equiv \frac{\sqrt{\pi} r_+}{8\sqrt\theta \gamma\left(\,3/2\ , r^2_+/4\theta\,\right) } \left(\, \frac{2}{\sqrt\pi}
\gamma\left(\,3/2\ , r^2_+/4\theta\,\right)-1\,\right)
\end{eqnarray}
$T_H$ is defined for any $r_+\ge r_0$.
The profile of $T_H(r_+)$ resembles that found in  \cite{Nicolini05} and implies a ``SCRAM phase'' at the terminal phase of the evaporation, namely a cooling
down to an asymptotic absolute zero configuration. The correction due to the prefactor $e^{\Phi}$ plays a
non-trivial role only near the peak of the temperature by lowering of the
maximum value, as it can be seen in figure (\ref{temp}). This is in  agreement with the choice of the equation (\ref{steqn}) and the consequent violations of WEC in a
region inside the event horizon $r_+$ \cite{Visser:1992qh}.\\
Once the Hawking temperature is known let us look at the area/entropy law.
From the first law of black hole thermodynamics, i.e.  $dM = T_H dS$, we can write the infinitesimal
entropy variation  $dS$  in terms of the  effective potential $U\left(\, r_+\,\right)$:

\begin{equation}
dS =\frac{1}{T_H} \frac{\partial U}{\partial r_+} dr_+ \label{ds}
\end{equation}

In order to integrate (\ref{ds}) in the correct way, we must take into account that the extremal, zero temperature, black hole configuration has zero thermodynamical entropy. Thus, the
integration  range starts from the radius $r_0$ of the degenerate horizon, and runs up to a generic
radius $r_+> r_0$.  With this choice of the integration range we find

\begin{equation}
 S=\frac{\pi^{3/2}}{2}\left[\, \frac{r_+^2}{\gamma\left(\,3/2\ , r^2_+/4\theta\,\right)} e^{-\Phi(r_+)}
-\frac{r_0^2}{\gamma\left(\,3/2\ , r^2_0/4\theta\,\right)} e^{-\Phi(r_0)}\,\right]-\Delta S
\label{entropia}
\end{equation}
where,
\begin{equation}
\Delta S\equiv  \frac{\pi^{3/2}}{2}\int_{r_0}^{r_+} dr r^2 \frac{d}{dr}\left[\,
 \frac{e^{-\Phi(r)}}{ \gamma\left(\,3/2\ , r^2/4\theta\,\right)}\,\right]
\end{equation}
The first term in (\ref{entropia}) differs from the corresponding quantity in the
clean case only for the presence of the exponential factors.  On the other hand,
the correction $\Delta S$ is characteristic of the dirty black hole, and vanishes
in the large distance limit. Indeed, for $r_+\ , r_0>> \sqrt\theta$ we recover
 the classical Area/Entropy law
\begin{equation}
S\to \frac{1}{4}\left(\, A_+ - A_0\,\right) \label{1/4}
\end{equation}
At first glance, the reader could question why the celebrated Bekenstein-Hawking
   (\ref{1/4}) results to be valid  only asymptotically, and not for any value of $r_+$.
 The key observation is that nowadays everybody is  taking for granted that the classical
 entropy of any black hole ``is'' one forth of the area of event horizon, measured
 in Planck units, up to eventual logarithmic corrections from unspecified quantum gravity effects.
 Thus, nobody cares anymore to ``recover'' the area law from basic principles. Rather, the main interest
 is to match the classical result with the statistical interpretation of entropy in terms
 of black hole quantum micro-states.  We would like to notice that the
 consistency between the geometric and thermodynamical definition of the black hole entropy requires that in any case it must be possible to \textit{derive} the Area/Entropy law  \textit{from} the first law (\ref{ds}), as we did
above, and to assume \textit{a priori} it is valid.  It is a common feature of regular black holes to display a more complex relation between horizon area and entropy, rather than a simple proportionality law.  This effect can be traced back, once again, to the presence of a minimal length or, equivalently, to the
 granular nature of the ``quantum'' spacetime. This effect, together with the existence of a finite
 leftover by the Hawking process, hints to a possible resolution of the information paradox
 worth of a more in-depth investigation.

\section{Conclusions}
In this paper we  extended our previous investigation of black hole solutions of
Einstein equations with a Gaussian source. Gaussian distributions are widely
used in physics and it is surprising that nobody considered before the gravitational
effects of this kind of sources. From our vantage point, we used this distribution
to model the physical effects of short-distance fluctuations of noncommutative coordinates.
Generally covariant divergence free condition on the energy-momentum tensor in the Einstein
equations allows various kind of physically acceptable conditions between the energy density and radial pressure. In our
case the component $T^{00}$ is assigned and the other ones must be determined in a consistent
way. After recalling the wormhole spacetime geometry
of the type conjectured in \cite{Garattini:2008xz}, we investigated a new dirty black hole type solution.
The novelty of this solution with respect to the corresponding ``clean'' case is
displayed in figure (\ref{WSEC}). While in the original noncommutative black hole \cite{Nicolini06}
only the strong energy condition is violated, in the dirty case weak energy conditions are
broken as well. The former violation removes the curvature singularity in the origin,
the latter decreases the Hawking temperature in agreement with the general result by Visser
\cite{Visser:1992qh}.

If LHC will turn to be an effective black hole factory, all the regular objects we have
investigated so far are expected to contribute to the cross section production.

\ack
P.N.~was partially supported by a CSU Fresno International Activities Grant. P.N. is supported by the Helmholtz International Center for FAIR within the
framework of the LOEWE program (Landesoffensive zur Entwicklung Wissenschaftlich-\"{O}konomischer Exzellenz) launched by the State of Hesse. P.N. would like to thank both the Max Plank Institute for Gravitational
Physics (Albert Einstein Institute), Golm, Potsdam, Germany and the CERN Theory Division for the kind hospitality during the final period of work on this project.


\section*{References}
\begin{thebibliography}{10}

\bibitem{Nicolini06}
Nicolini P, Smailagic A and Spallucci E 2006
\PL  B {\bf 632} 547


\bibitem{Rizzo06}
Rizzo T G 2006 {\it J. High Energy Phys.} JHEP09(2006)021
%

\bibitem{Ansoldi07}
Ansoldi S, Nicolini P, Smailagic A and Spallucci E 2007 \PL  B {\bf 645} 261


\bibitem{Spallucci:2008ez}
Spallucci E, Smailagic A and Nicolini P 2009
  \PL B {\bf 670} 449 


\bibitem{Casadio08}
Casadio R and Nicolini P 2008 {\it J. High Energy Phys.}  JHEP11(2008)072 

\bibitem{Ansoldi08}
  Ansoldi S Spherical black holes with regular center: a review of existing models
  including a recent realization with Gaussian sources {\em Preprint}
  arXiv:0802.0330 [gr-qc]

\bibitem{Nicolini08}
Nicolini P 2009
  {\it Int.\ J.\ Mod.\ Phys.\ } A {\bf 24} 1229 
  
\bibitem{Arraut:2009an}
  Arraut I, Batic D and Nowakowski M 2009 A non commutative model for a mini black hole {\it Preprint}
  arXiv:0902.3481 [gr-qc]

\bibitem{Smailagic:2001qe}
  Smailagic A and Spallucci E 2002
  \PR  D {\bf 65} 107701 

\bibitem{Smailagic:2002sp}
  Smailagic A and Spallucci E 2002
  \JPA  {\bf 35} L363 
  
\bibitem{Smailagic:2003yb}
  Smailagic A and Spallucci E 2003
  \JPA  {\bf 36} L467
  
\bibitem{Smailagic:2003rp}
  Smailagic A and Spallucci E 2003
  \JPA  {\bf 36} L517

\bibitem{Smailagic:2004yy}
 Smailagic A and Spallucci A 2004
  \JPA  {\bf 37} 1
  [Erratum-ibid.\  A {\bf 37} 7169]

\bibitem{Nicolini:2005de}
  Nicolini P, Smailagic A and Spallucci E 2005 The fate of radiating black holes in noncommutative geometry {\em Preprint}
  arXiv:hep-th/0507226

\bibitem{Nicolini05}
Nicolini P 2005 \JPA  {\bf 38} L631
%

\bibitem{Spallucci06}
  Spallucci E, Smailagic A and Nicolini P 2006
  \PR  D {\bf 73} 084004 
  
\bibitem{Banerjee:2009gr}
  Banerjee R, Chakraborty B, Ghosh S, Mukherjee P and Samanta S 2009
  Found.\ Phys.\  {\bf 39}, 1297 

\bibitem{Nicolini:2009dr}
Nicolini P and Rinaldi M 2009 A minimal length versus the Unruh effect {\em Preprint}
  arXiv:0910.2860 [hep-th]
  
  \bibitem{Sushkov:2005kj}
  Sushkov S V 2005
  \PR  D {\bf 71} 043520 
  
  \bibitem{Morris:1988cz}
  Morris M S and Thorne K S 1988
{\it  Am.\ J.\ Phys.\ } {\bf 56} 395



\bibitem{Visser:1992qh}
  Visser M 1992
  \PR  D {\bf 46} 2445 
  
\bibitem{Garattini:2008xz}
  Garattini R and Lobo F S N 2009
  \PL B {\bf 671} 146 


\bibitem{Gibbons:1987ps}
Gibbons  G W and Maeda K I 1988
  \NP  B {\bf 298} 741


\bibitem{Ichinose:1989vb}
Ichinose I and Yamazaki H 1989
{\it  Mod.\ Phys.\ Lett.\ } A {\bf 4} 1509 

\bibitem{Yamazaki:1992pm}
Yamazaki H and Ichinose I 1992
  \CQG  {\bf 9} 257 
  
\bibitem{Garfinkle:1990qj}
Garfinkle D, Horowitz G T and Strominger A 1992
  \PR  D {\bf 43} 3140 
  [Erratum-ibid.\  D {\bf 45} 3888]

\bibitem{Allen:1989kc}
  Allen T J, Bowick M J  and Lahiri A 1990
  \PL  B {\bf 237} 47 

\bibitem{Campbell:1991rz}
  Campbell B A, Kaloper N and Olive K A 1991
  \PL  B {\bf 263} 364 

\bibitem{Lee:1991jw}
Lee K M and Weinberg E J 1991
  \PR  D {\bf 44} 3159

\bibitem{Dowker:1991qe}
Dowker F, Gregory R and Traschen J H 1992
  \PR  D {\bf 45} 2762 

\bibitem{Shapere:1991ta}
  A.~D.~ShapereA D, Trivedi S and Wilczek F 1991
 {\it Mod.\ Phys.\ Lett.\ } A {\bf 6} 2677 

\bibitem{Galtsov:1989ip}
  Galtsov D V and Ershov A A 1989
  \PL  A {\bf 138} 160

\bibitem{Straumann:1990as}
  Straumann N and Zhou Z H 1990
  \PL  B {\bf 243} 33 
  
  \bibitem{Bizon:1990sr}
  Bizon P 1990
  \PRL  {\bf 64} 2844

  \bibitem{Bizon:1991nt}
Bizon P and Wald R M 1991
  \PL  B {\bf 267} 173 

 \bibitem{Lahiri:1992yz}
  Lahiri A 1992
  \PL  B {\bf 297} 248 


\end{thebibliography}
\end{document}